\documentclass[apl,superscriptaddress,twocolumn,preprintnumbers,a4,showpacs,floatfix]{revtex4-1}
\usepackage[T1]{fontenc}
\usepackage[english]{babel}

\usepackage[sort&compress]{natbib}

\usepackage{color}

\usepackage[latin1]{inputenc} % tämän pitäisi toimia oletuksena ainakin Texnic-centerissä ja TeXMakerissa
\usepackage{mathtools}
\usepackage{amsmath}
\usepackage{amsfonts}
\usepackage{amssymb} % Tarvitaan mm. lukualueiden symbolien kirjoittamiseen komennolla \mathbb{}.
\usepackage{amsthm} % mm. tarjoaa lause- ja todistusympäristöt.
\usepackage{graphicx} % Kuvien lisäämistä varten
\usepackage{verbatim} % Tarjoaa mm. kommenttiympäristön.
\usepackage{enumerate} % Lisää vaihtoehtoja numeroitujen listojen muotoiluun.
\usepackage{setspace} % Rivivälin helppoa muuttamista varten.
\usepackage{booktabs} % Tarjoaa \toprule-, \midrule- ja \bottomrule-komennot taulukkoja varten

\newcommand{\be}{\begin{equation}}
\newcommand{\ee}{\end{equation}}

\def\bear{\begin{eqnarray}}
\def\eear{\end{eqnarray}}

\begin{document}

%Muutetaan riviväliksi 1,5 tiettyjä paikkoja, kuten kuvatekstejä ja alaviitteitä, lukuun ottamatta.
%\onehalfspacing
\setcounter{page}{1}

% Limits of stable bents for graphene nanoribbons on substrate
% Limits of stable bents for supported graphene nanoribbons
% Deformation limits of bent and supported graphene nanoribbons
%\title{Limits of stable bents for supported graphene nanoribbons}
\title{Limits of stability in supported graphene nanoribbons subject to bending}

\author{Topi Korhonen}
\author{Pekka Koskinen}
\email[email:]{pekka.koskinen@iki.fi}
\address{NanoScience Center, Department of Physics, University of Jyvaskyla, 40014 Jyv\"askyl\"a, Finland}

% 68.65.Pq 	Graphene films
% 62.25.-g 	Mechanical properties of nanoscale systems
%68.35.Gy 	Mechanical properties; surface strains
%68.55.-a 	Thin film structure and morphology
%68.60.-p 	Physical properties of thin films, nonelectronic
%68.60.Bs 	Mechanical and acoustical properties 

%68.65.-k 	Low-dimensional, mesoscopic, nanoscale and other related systems: structure and nonelectronic properties
%61.46.-w 	Structure of nanoscale materials

\begin{abstract}
Graphene nanoribbons are prone to in-plane bending even when supported on flat substrates. However, the amount of bending that ribbons can stably withstand remains poorly known. Here, by using molecular dynamics simulations, we study the stability limits of $0.5-1.9$~nm wide armchair and zigzag graphene nanoribbons subject to bending. We observe that the limits for maximum stable curvatures are below $\sim 10$~deg/nm, in case the bending is externally forced and the limit is caused by buckling instability. Furthermore, it turns out that the limits for maximum stable curvatures are also below $\sim 10$~deg/nm, in case the bending is not forced and the limit arises only from the corrugated potential energy landscape due to the substrate. Both of the stability limits lower rapidly when ribbons widen. These results agree with recent experiments and can be understood by means of transparent elasticity models.
\end{abstract}

\maketitle
Today graphene nanoribbons can be fabricated at atomic precision, but only in the presence of a stabilizing substrate.\cite{Atomically_precise_bottom_up_fabrication_of_graphene_nanoribbons_nat}
The substrate stabilizes flimsy ribbons and suppresses their tendency to twist, fold and ripple.\cite{Intrinsic_ripples_in_graphene_nat, Unusual_ultra-low-frequency_fluctuations_in_freestanding_graphene_natCom,bets_NR_09,Ramasubramaniam2012,kit_PRB_12} However, even substrates cannot fully prevent all deformations,
most of which induce mechanical strains that alter ribbons' electronic properties.\cite{Chemically_Derived_Ultrasmooth_Graphene_Nanoribbon_Semiconductors_sci,effects_of_strain_on_electronic_properties_of_graphene_prb, Strain_Effect_on_the_Electronic_Properties_of_Single_Layer_and_Bilayer_Graphene_jpc} Actually, such strain engineering of electronic properties is gaining popularity, whereby detailed knowledge of mechanical stability limits is becoming increasingly valuable.\cite{strain_engineering_of_graphenes_electronic_structure_prl}

Mechanical strain can be created for example by lattice mismatch, by impurities and lattice defects, and by the fabrication process itself.\cite{periodically_rippled_graphene_prl} Compressive strain, in particular, is often limited by buckling instability. For uniaxial compression buckling has been observed in experiments at $0.5$~\%\ strain and in simulations at $0.8$~\%\ strain.\cite{compression_behaviour_of_single_layer_graphene_acs,mechanical_properties_of_monolayer_graphene_under_tensile_and_compressive_loading_pe} In graphene nanoribbons, however, the most pertinent deformation is not uniaxial compression but bending. Yet, the mechanical stability limits of supported ribbons subject to bending remain unexplored. In this letter, therefore, we aimed to address two fundamental questions: How much can a graphene nanoribbon of given width bend on a given substrate until it buckles? And, to what extent can it remain bent due to the corrugation potential energy of the substrate alone, without external forcing? As it will turn out, both of these questions could be answered by transparent modeling.

\begin{figure}[b!]
\centering
\includegraphics[width=.5\textwidth]{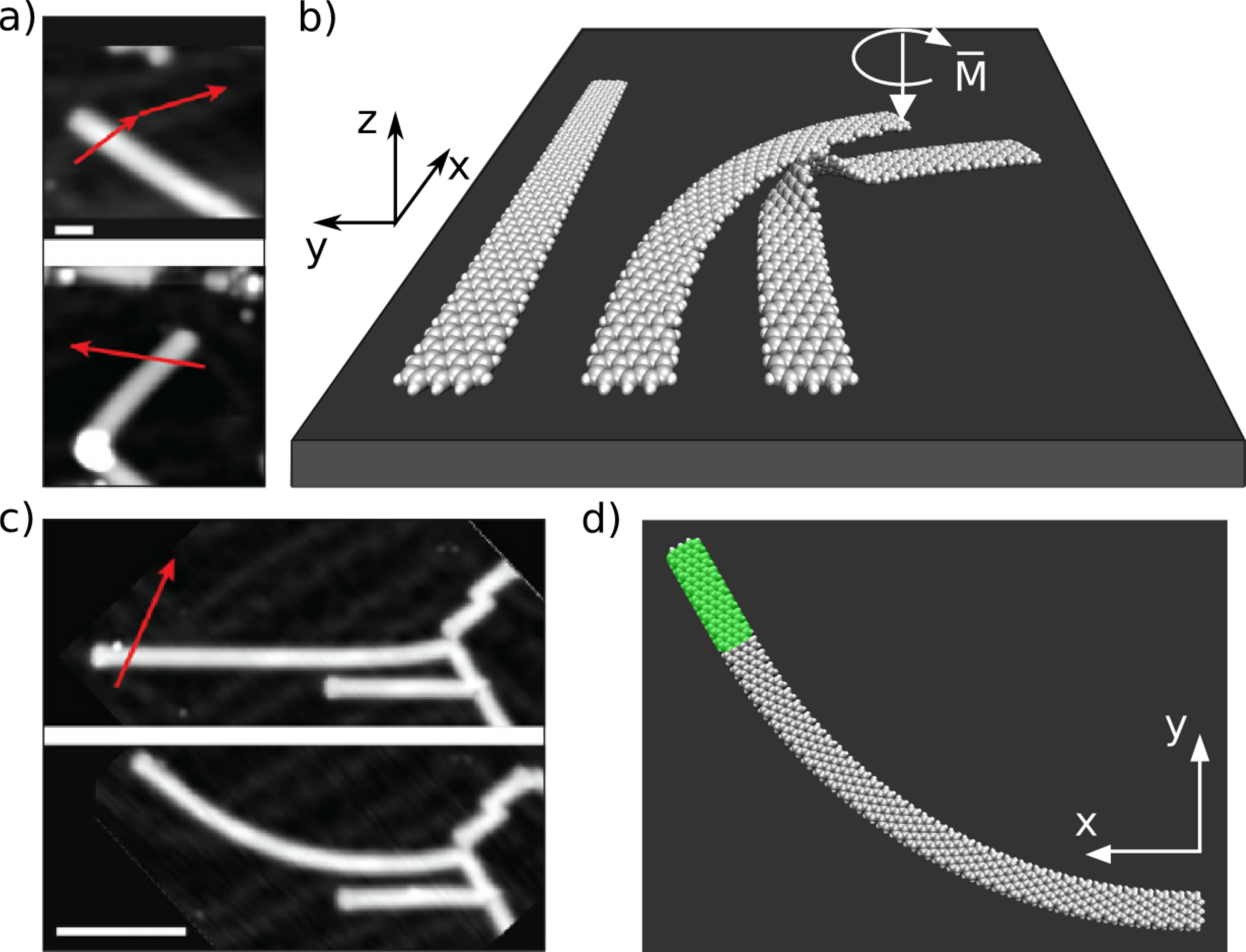}
\caption{(color online) 7-armchair graphene nanoribbons subject to bending. (a) In experiments bending was controlled by the tip of an atomic force microscope, whose movements are denoted by arrows. Buckling is seen as the bright kink. (b) In simulations ribbons were bent by fixing their front ends and by turning their tail ends. The rightmost geometry shows the buckled geometry. (c) Maximum curvature without external forcing. After manipulation the ribbon remained bent by the substrate corrugations alone. Scale bar, $10$~nm. (d) In the simulations one end of the ribbon (green tail) was pinned to (set in registry with) the substrate while the other end was turned to the maximum stable curvature beyond which the entire ribbon started sliding. The experimental figures in panels (a) and (c) are reproduced from Ref.~\onlinecite{bending_and_buckling_iop} by Creative Commons Attribution licence; image ordering has been changed.} 
\label{pic:general}
\end{figure}

Our simulations were closely related to the recent experiments of van der Lit \emph{et al.} in Ref.~\onlinecite{bending_and_buckling_iop} (Fig. 1). There an atomically precise 7-armchair graphene nanoribbon was bent at low temperature on Au(111) surface by an atomic force microscope (AFM) tip. Under forced bending and above certain maximum curvature the ribbon was observed to buckle off the substrate (Figs.~1a). Furthermore, ribbon was observed to withstand certain maximum curvature, presumably due to the lateral energy corrugations arising solely from the substrate interactions (Figs.~1c). 

\begin{figure}[t!]
	\centering
	\includegraphics[width=.4\textwidth]{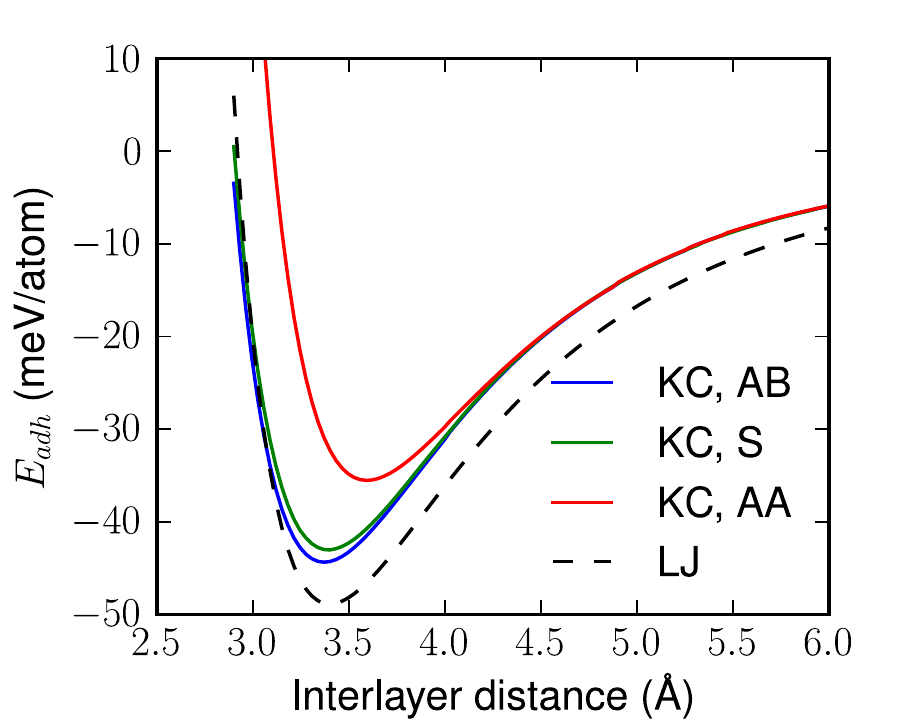}
\caption{(color online) Ribbon's adhesion energy per atom as a function of distance from the substrate. Under superlubric conditions surface adhesion is modeled by laterally homogeneous Lennard-Jones (LJ) potential. Under conditions where registry effects are important, the adhesion is modeled by Kolmogorov-Crespi (KC) potential, which models energy corrugations by making the energy minimum registry-dependent (shown with adhesion curves for AA, AB, and saddle (S) point configurations).\cite{KC}}
\label{fig:adhesion}
\end{figure}

To investigate the buckling instability in more detail, we simulated ribbons subject to forced bending (Fig.~1b). We simulated hydrogen-passivated $N$-armchair ($N=5,\;7,\;9,\;11,$ and $13$) and $N$-zigzag ($N=4,\;6,\;8,$ and $10$) graphene nanoribbons of widths $w\approx 0.5-1.9$~nm and lengths given by $1/10$ aspect ratio. The C-C, C-H, and H-H interactions were modeled by the empirical reactive bond-order potential REBO.\cite{REBO} The ribbons were initially relaxed on a model Au substrate, which assumed an interaction with the ribbon described by a $z$-dependent potential with $20$~meV/\AA$^2$ adhesion, $3.4$~\AA\ equilibrium distance, and a functional form suggested by the Lennard-Jones 12-6 potential (Fig.\ref{fig:adhesion}).\cite{vdw_grapheneAu,van_der_Waals_Bonding_in_Layered_Compounds_prl,Koskinen2014} This substrate model ignores lateral energy corrugation, but it is expected to be a good approximation, because graphene nanoribbons that are out of registry with respect to the Au(111) substrate have been shown not to experience any lateral forces, and thus to exhibit superlibricity.\cite{superlubric_acs, Kawai2016}

The supported ribbons were simulated by the LAMMPS code, using $1$~fs time step and Langevin thermostat at $10$~K temperature and $5$~ps damping time.\cite{LAMMPS} First the ribbons were thermalized on the model substrate. Then they were gradually bent by fixing one end and slowly (quasi-statically) turning the other end while simultaneously allowing its free movement in the plane (Fig.~1b). At a later instant the turning direction was reversed, and the simulation terminated with straight ribbons.

At the initial stages of the simulations the bending was smooth and the ribbons remained adhered to the substrate. Here we quantify the amount of bending both by the in-plane curvature $\kappa=1/R$, where $R$ is the radius of curvature, and by the dimensionless curvature $\Theta = \kappa w/2$, which also equals the absolute amount of strain at the ribbon edges. Using straightforward continuum elasticity theory, the elastic energy during this initial stage is
\bear
\label{eq:potE}
E_\text{bend}(\Theta) = (1/6)k w l \Theta^2 [1 - 2 \tau/(k w)]^2,
\eear
where $w$ is ribbon width, $l$ is ribbon length, $k = 19$~eV/\AA$^2$ is graphene's in-plane modulus, and $\tau$ is the stress at the passivated armchair ($\tau_{ac} = -1.5$~eV/\AA) or zigzag edges ($\tau_{zz} = -0.2$~eV/\AA), as given by the REBO potential.\cite{edge_elastic_properties_apl} 
Eq.~(\ref{eq:potE}) gives the elastic energy below $\Theta\lesssim 3$~\%\ at fair accuracy (Fig. 3a).

During this initial stage we observed weak ripples at the inner edges of the ac-ribbons. Ripples were notable up- and down-displacements of alternating armchair units and observable along the entire ribbon. They have been observed also in straight ribbons where they have been attributed to chemically induced edge stress; here the edge stress was created mostly by the bent geometry itself.\cite{quantum_edge_stress_prl, edge_stress_induced_warping_prl} When curvature increased, the rippling amplitude increased, but wavelength remained fixed. These ripples were observed only for the ac-ribbons as zz-ribbons remained almost completely flat prior to bucling.  

When the increasing curvature reached a critical limit, the in-plane stress finally became unbearable and the ribbon suddenly buckled (Fig. 1c). Buckling allowed two parts of the ribbon to straighten, which released in-plane elastic energy, although at the expense of lost adhesion and increased out-of-plane bending energy. Buckling occurred later for narrow ribbons than for wide ribbons. The events during the bending-straightening simulations are best gauged through the maximum height of the ribbon above the substrate (Fig.~3b). Initially the buckle was formed at $\Theta_{b'}$, but upon straightening it remained stable also for curvatures $\Theta<\Theta_{b'}$ so that when the ribbon finally unbuckled at $\Theta_b$, roughly half the buckling curvature, the result was a notable hysteresis. The buckling-unbuckling process was reversible; plastic deformations did not occur. These observations are in agreement with experiments that also showed the restoring of the initial geometry. In particular, for $N=7$ ac-ribbon the buckling occurred in experiments at curvature of $4$~deg/nm, in reasonable agreement with the computational curvature of $6$~deg/nm.\cite{bending_and_buckling_iop} Note that it is justifiable to compare experiments only to the smaller curvature $\Theta_b$, because in macroscopic time scales random perturbations help drive the system toward buckled geometry already at smaller curvatures.

\begin{figure}[tb!]
	\centering
	\includegraphics[width=.45\textwidth]{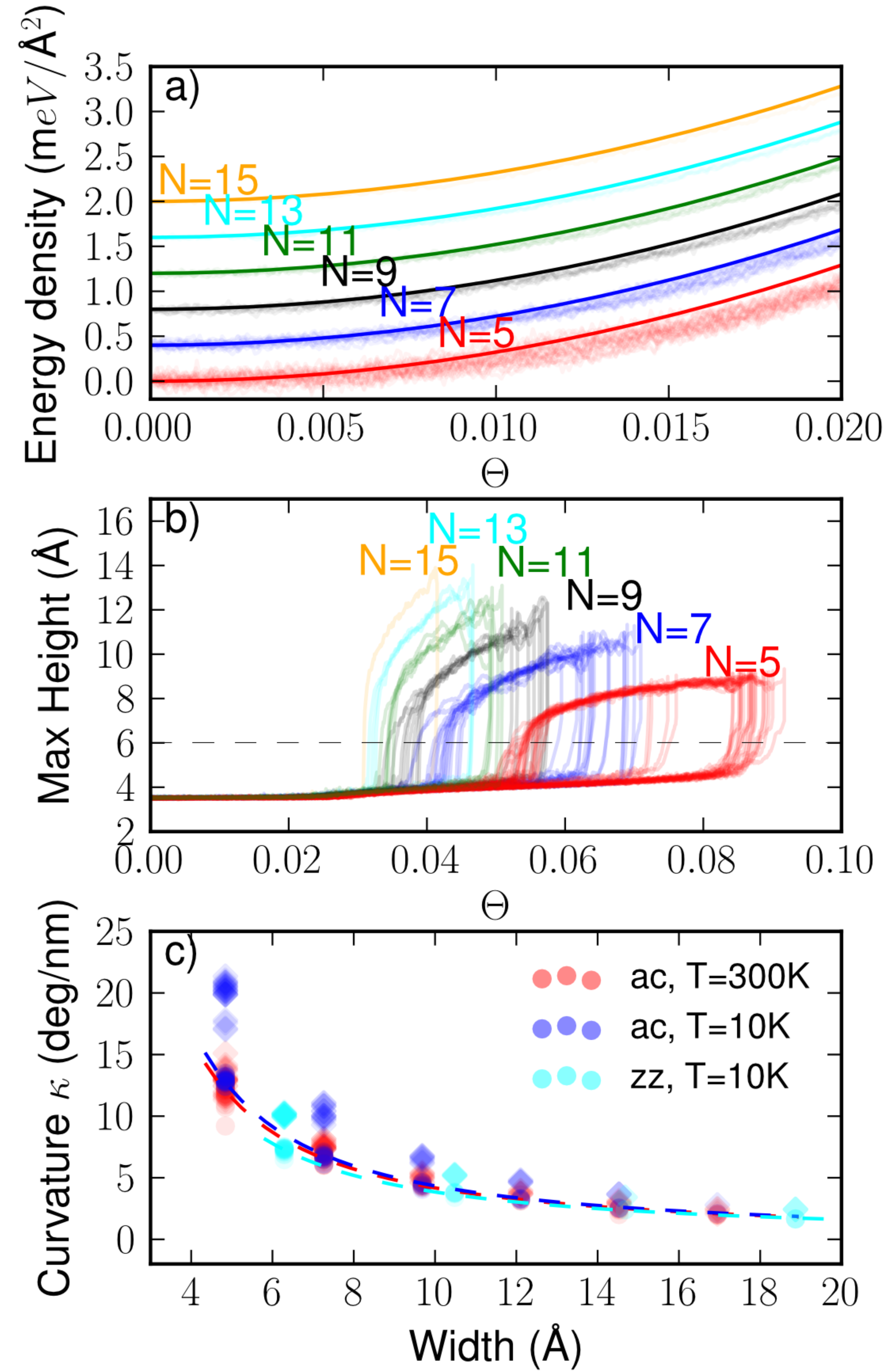}
\caption{(color online) Trends in buckling instabilities. (a) Simulated elastic energy densities (thin wiggly lines) compared with the elastic model of Eq.~(\ref{eq:potE}) (thick solid lines) for ac-ribbons of different widths. Curves are offset for clarity. (b) The maximum height of the ac-ribbons above the substrate. The bending and straightening simulations show hysteresis in the buckling: buckling requires larger curvature than unbuckling. Dotted line is the buckling threshold. (c) Buckling and unbuckling curvatures for different ribbons and temperatures as defined by the threshold in panel b.}
\label{fig:3}
\end{figure}

To understand the general width-dependence in the buckling (Fig.~\ref{fig:3}c), let us develop a model that accounts for the in-plain strain, out-of-plain bending, and substrate adhesion energies. In the model the ribbon is treated as two aligned narrow strands that represent the compressed and stretched halves of the ribbon. The aligned strands are next to the neutral line and separated by $w_{eff} = \alpha w$, where the width-dependent parameter $\alpha$ ($\lesssim 1$) is later fitted to account for the averaging. Upon buckling the outer strand remains flat but the height profile of the inner strand acquires the form $y(l) = A \sin^2(l/\lambda \pi)$ ($0\leq l \leq \lambda$), where $A$ is the buckling amplitude and $l$ is the distance measured along the strand. This profile decreases the strand length by $\Delta l = \pi^2A^2/(4 \lambda)$ and thereby relieves the compressive strain energy at the inner edge by $wk\Theta \Delta l/2$ and the tensile strain energy at the outer edge by the same amount. This approach is similar to that in Ref.~\onlinecite{bending_induced_delam_cm}. Adding this strain energy release to the loss in Lennard-Jones energy ($\int w/2[V_{LJ}(y)-V_{LJ}(u)]\text{d}l$) and the out of plane bending energy associated with the height profile ($\int \frac{w}{4} D y''^2(l)\text{d}l$), the energy difference between purely bent and buckled ribbon becomes
\be
\label{eq:pin}
\Delta E(\Theta) = A^2 \frac{w}{2} \left[-k \frac{\pi^2 A^2}{2 \lambda}\Theta \alpha + \frac{15}{2}\frac{\epsilon_{vdw}\lambda}{\sigma^2} + \frac{D \pi^4}{\lambda^3} \right].
\ee    
Here $\epsilon_{vdw}$ is the adhesion energy per unit area, $\sigma$ is the interlayer distance, and $D = 1.0$~eV is graphene's bending modulus.\cite{bending_induced_delam_cm, Lattice_Dynamics_of_Pyrolytic_Graphite_prb} Buckling occurs when the first term becomes large enough due to the increasing curvature so that $\Delta E(\Theta_b) = 0$. The energy of the buckled geometry is further minimized by $\partial \Delta E/\partial \lambda |_{\lambda=\lambda_b}=0$. Solving these equations yields $\lambda_b = 9$~\AA\ and
\be
\label{eq:theta0}
\Theta_b(T) = 2/(k \sigma \alpha)\sqrt{30 \epsilon_{vdw}D} \approx 0.023 \times\alpha^{-1}.
\ee
Fit to the simulations gives $\alpha_i = 1/(\beta_i w^{-1}+1)$, where $\beta_{ac}=7$~\AA\ and $\beta_{zz} = 5$~\AA, which provide a good agreement with the simulated buckling curvatures (Fig.~\ref{fig:3}c). The fit is physically meaningful and obeys the consistency requirement $\alpha\lesssim 1$. The validity of the model is probably limited for ribbon widths below few nanometers, although $\Theta_b=2.3$~\%\ is a reasonable limit for very wide ribbons, too.

While our simulations included ribbons only with hydrogen-passivated zigzag and armchair edges, also other edges with other passivations or edge reconstructions are possible.\cite{Koskinen1, Koskinen2, Suenaga2010, Dresselhaus} Especially in free-standing graphene the edges may create sizable corrugations.\cite{dist1, edge_stress_induced_warping_prl} On substrates these corrugations diminish in magnitude, but do not vanish completely.\cite{Reddy} However, here the edge stresses are small due to hydrogen passivation and the lateral stresses due to bending are so large that the effect of edge stress is fairly small. This is suggested already by the quantitatively similar buckling behavior in zigzag and armchair ribbons (Fig.~\ref{fig:3}c).

% edges:
% 1 P. Koskinen, S. Malola, and H. Häkkinen, Phys. Rev. Lett. 101, 115502 (2008).
% 1 P. Koskinen, S. Malola, and H. Häkkinen, Phys. Rev. B 80, 073401 (2009).
% 1 K. Suenaga and M. Koshino, Nature 468, 1088 (2010).
% 1 X.J. dn J. Campos-Delgado, M. Terrones, V. Meunier, and M.S. Dresselhaus, Nanoscale 3, 86 (2011).

% distortions:
%1 V.B. Shenoy, C.D. Reddy, and Y.-W. Zhang, ACS Nano 4, 4840 (2010).
% 1 V. Shenoy, C. Reddy, A. Ramasubramaniam, and Y. Zhang, Phys. Rev. Lett. 101, 245501 (2008).

% reddy:
%Influence of substrate on edge rippling in graphene sheets C D Reddy1, Yong-Wei Zhang1 and V B Shenoy2 --> TAMA PITAISI SAADA JA LUKEA
%@article{linear_elastic_graphene3,
%  author={C D Reddy and Yong-Wei Zhang and V B Shenoy},
%  title={Influence of substrate on edge rippling in graphene sheets},
%  journal={Modelling and Simulation in Materials Science and Engineering},
%  volume={19},
%  number={5},
%  pages={054007},
%  url={http://stacks.iop.org/0965-0393/19/i=5/a=054007},
%  year={2011},
%}

For completeness, we repeated buckling simulations for armchair ribbons also at room temperature. As the main result, the effect of temperature was to reduce the hysteresis and initiate buckling at slightly smaller curvatures (Fig. \ref{fig:3}c). On average, however, the buckling occurred around the same curvature as described by the model fitted at low temperature.

% TOINEN SIMULI
In the next set of simulations, we investigated the limits of maximal curvature in armchair ribbons allowed by the substrate energy corrugation alone. In these simulations we chose to place the ribbons on a graphene substrate modeled by the Kolmogorov-Crespi (KC) registry-dependent interlayer potential.\cite{KC} This model substrate was obviously different from the Au(111) substrate in the experiments, but our choice was a necessary compromise for a feasible substrate model with a realistic energy corrugation. Namely, the frequently used Lennard-Jones potential typically yields an order of magnitude too low energy corrugation for sliding, and proper registry-dependent potentials for graphene and Au(111) are missing.\cite{reguzzoni_PRB_12} Nevertheless, the ribbon adhesions for both Au and graphene substrates are similar, so the KC potential was an attempt to combine a well-defined substrate model with a realistic corrugation energy landscape.

In these simulations one end of the ribbon was first appended by a tail of length $L_t$ that was pinned to the substrate by setting it in full registry (Fig.~1f). The other end was then gradually turned until the maximum stable curvature beyond which the pinning was released and the tail started sliding, causing straightening of the ribbon. The ribbon was considered stable at given $L_t$ and $\kappa$ if it remained in place for $20$~ps, although it was evident already within few ps whether the curvature was stable or not. The maximum curvature limits were then searched for each ribbon width with several tail lengths.
   
\begin{figure}[t!]
\centering
\includegraphics[width=.4\textwidth]{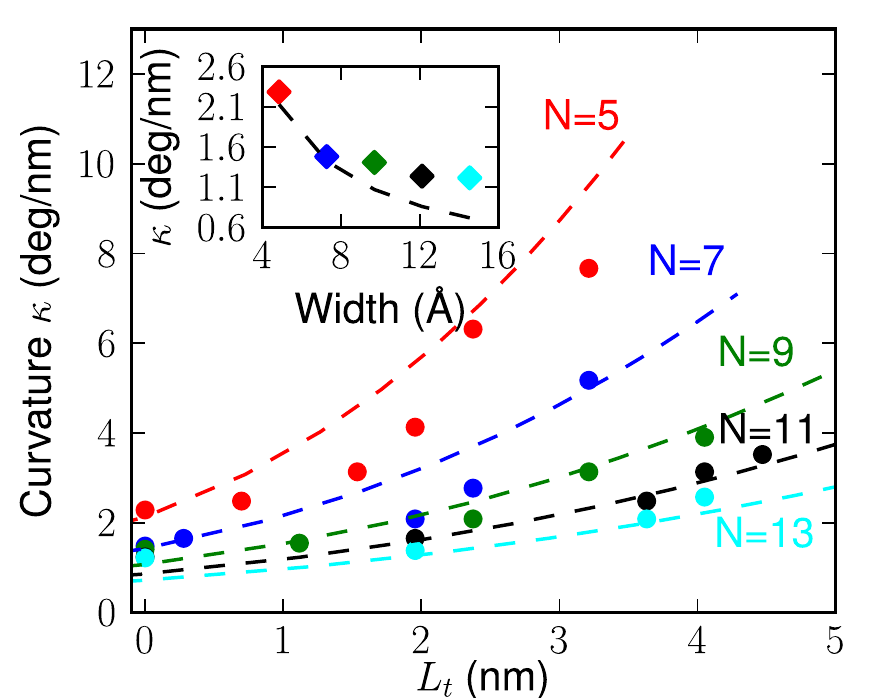}
\caption{(color online) Maximum stable in-plane curvatures for different ribbons as a function of the added tail length $L_t$. Dashed lines are the model estimates from Eq.~\ref{eq:moments}. Inset: maximum curvature limit as a function of ribbon width at $L_t=0$.}
\label{pic:general}	
\end{figure}

Simulations show that narrow ribbons withstand higher curvatures than wide ribbons and that maximum curvatures increase when the tail lengths increase (Fig.~4). It is notable that certain finite curvatures can be achieved even in the absence of any added tail (inset of Fig.~4). This occurs because also ribbon's end is close to registry and not yet subject to superlubric behavior. By geometry considerations we therefore approximate that the length $L_{t'} = \sqrt{2Ra + a^2}$ close to the end of the ribbon is still pinned to the substrate, where $a$ is a length scale for the tolerance in a lateral displacement that is still considered to be in registry. Thus, the total length of the substrate-pinned ribbon at the end equals $L_\text{pin}=L_{t'}+L_t$. This assumption serves as a starting point for a model for the maximum curvature limit. In the model we consider the pinned part to be subject to a bending moment $kw^2\Theta/6$ imposed by the unpinned part. This moment must not exceed a maximum value, lest the pinned part starts to slide. At the maximum the bending moment equals the maximum allowed moment, or
\be
\label{eq:moments}
\frac{1}{6}kw^2\Theta = \int_\text{pinned} r\times( f\text{d}\mathcal{A}), 
\ee 
%\be
%\label{eq:moments}
%\frac{1}{6}kw^2\Theta = f\int_0^L \int_{-w/2}^{w/2} \sqrt{x^2 + y^2} \text{d}x\text{d}y, 
%\ee  
where $f$ is the maximum force per unit area during sliding, averaged over all sliding directions. The integration is over the pinned part of length $L_\text{pin}$ and $r$ is the distance to its center of mass. Fitting the force parameter $f$ with a chosen tolerance $a=0.7$~\AA \ to simulation data yields $f=0.7$~meV/\AA$^3$.
%Fitting the existing two parameters to simulation data yields $a=0.7$~\AA\ and $f=0.7$~meV/\AA$^3$. 
The maximal force per unit area for sliding in an armchair direction is $f_{max}=2.3$~meV/\AA$^3$, which confirms the physical interpretation of the fit ($f\approx 0.3\times f_{max}$).\cite{Korhonen2015} 

Upon inserting these parameters into the model Eq.~(\ref{eq:moments}), the trends in maximum curvatures get reproduced surprisingly well (Fig.~4). The model underestimates the maximum curvatures as compared to simulations, which is however not surprising given the highly discrete nature of the short-tail limit (inset of Fig.~4). The model predicts pinning at roughly constant edge strain of $\sim0.9\%$, but in simulations the allowed edge strain depends somewhat on ribbon width, changing as ribbons widen from $\sim0.9$~\%\ for $N=5$ to $\sim1.5$~\%\ for $N=13$. Such dependence may originate due to thermal fluctuations, which affect narrow and wide ribbons differently due to the different number of pinned atoms.

These simulations can be compared to the experimentally observed pinning in Ref.~\onlinecite{bending_and_buckling_iop}, although with caution. The energy corrugations for graphene ribbons on Au(111) and on graphene are probably different, but likely of similar magnitude due to the similarity of the adhesion itself.\cite{vdw_grapheneAu} To this end, note that the model in Eq.~(\ref{eq:moments}) suggests that the substrate affects the trends \emph{only} through the averaged parameter $f$. Thus, even though the symmetry in Au(111) differs from that in graphene, it is not unreasonable to expect that the results would correspond also to Au substrate, at least semi-quantitatively. Such correspondence is further supported by the rough agreement between the experimental ($2$~deg/nm for gold substrate) and simulated ($1.4$~deg/nm for model graphene substrate) maximum curvatures for a 7-armchair ribbon.\cite{bending_and_buckling_iop} At any rate, the parameter $f$ allows transferring the results to any other substrate, making the model highly versatile. 

%\cmtr{TEMPERATURE+PINNING DISCUSSION HERE? While the effect of temperature in buckling was small, in pinning it is large. In Fig.\ref{fig:adhesion} we see that in KC potential the energy corrugation per atom is only of the order of $9$~meV, which corresponds only to temperature of $T\sim 100$~K, much smaller than room temperature. Therefore around $T\sim 100$~K pinning does not occur anymore and jne jne. Jotakin sinnepain.}

While in buckling the effect of temperature was clearly small, in pinning its effect is more ambiguous. Although the energy corrugation per atom $\sim9$~meV corresponds only to the temperature of $T\sim 100$~K, the pinning still occurred also at room temperature, at least withing time scales accessible to the simulations ($20$~ps). The general tendency of an increased temperature was to modestly decrease the maximal pinning curvature, although the results became less clear. While at low temperatures the possible unpinning of the tail was fast ($\sim 1-2$~ps) and clear-cut, at high temperatures thermal fluctuations brought unambiguity by introducing more variations to the time scale of unpinning. Thus, reliable determination of structure stability would have required simulation times beyond reasonable limits, as also indicated by recently observed sliding phenomena.\cite{Wang2016}

To conclude, these simulations and the associated models provide transparent understanding for the stability limits in supported graphene nanoribbons subject to bending. Narrow $5$-, $7$-, $9$-, $11$-, and $13$-armchair ribbons require only minimal pinned parts to maintain curvatures around $1$~deg/nm (radius of curvature $R\approx 60$~nm). Although such curvatures are gentle, other studies have found them to cause predictable modifications in ribbons' electronic and optical properties. In particular, simulations in Ref. \onlinecite{Graphene_nanoribbons_subject_to_gentle_bends_prb} showed that the energy gap for $N$-armchair graphene nanoribbons change according to the expression
\begin{equation}
\Delta E_g(\Theta)=\frac{1}{2}(-1)^q\gamma \delta \Theta^2,
\label{eq:gap_change}
\end{equation}
where $q=\mod(N,3)$ (restricted to $q=0,\,1$) is the ribbon family, $\gamma=1.7$ describes bond anharmonicity that is relevant for bending-induced stretching, and $\delta=12$~eV is an electromechanical coupling constant related to gap changes during the stretching of straight ribbons. Combining Eq.(\ref{eq:gap_change}) with Eq.(\ref{eq:theta0}), the buckling-limited maximum energy gap change becomes directly 
\begin{equation}
|\Delta E_g^{max}(w)|=5.4\times(7\text{ \AA}\times w^{-1}+1)^2\text{ meV}.
\label{eq:gap_max}
\end{equation}
For the ribbons studied here this amounts from $23$~meV ($N=5$) to $10$~meV ($N=13$) gap changes. For wider ribbons the maximum gap change shrinks. In the case of pinning the maximum curvature depends on the tail length $L_t$, but it is always limited by Eq.(\ref{eq:theta0}), so with unconstrained bending Eq.(\ref{eq:gap_max}) gives the upper limit for gap changes.

Buckling, however, can modify the electronic properties even more than bending. Simulations showed that narrow ribbons remained flat above $4$~deg/nm curvatures ($R\approx 14$~nm), but stability was strongly width-dependent; ribbons wider than $1.5$~nm remained flat only below $2$~deg/nm ($R\gtrsim 29$~nm). The obtained stability limits thus provide guidelines to design experiments and to choose structures that would be stable enough for reliable device operation. Because the adhesion energies for most van der Waals bound, physisorbed two-dimensional materials are of similar magnitude, we expect the presented elastic models to have applicability for several other ribbon and substrate materials.\cite{van_der_Waals_Bonding_in_Layered_Compounds_prl} To this end, we propose that the stabilities of bent ribbons could even be used as a measurement technique to investigate the interaction between different nanoribbons and substrates. 

\emph{Acknowledgements:} We thank the Academy of Finland for funding (Projects No. 283103 \& 251216) and CSC - IT Center for Science in Finland for computer resources.

%\bibliographystyle{apsrev4-1}
%\bibliography{sS,pekkas}{}
%\bibliographystyle{plain}


\end{document}